\documentclass[twocolumn,showpacs,prl,aps, superscriptaddress]{revtex4}
\usepackage{amssymb}
\usepackage{amsmath}
\usepackage{graphicx}

\begin{document}

\title{Intervalley Scattering and Localization Behaviors of Spin-Valley Coupled Dirac Fermions}

\author{Hai-Zhou Lu}
\affiliation{Department of Physics and Centre of Theoretical and Computational Physics,
The University of Hong Kong, Pokfulam Road, Hong Kong, China}

\author{Wang Yao}
\affiliation{Department of Physics and Centre of Theoretical and Computational Physics,
The University of Hong Kong, Pokfulam Road, Hong Kong, China}

\author{Di Xiao}
\affiliation{Department of Physics, Carnegie Mellon University, Pittsburgh, Pennsylvania 15213, USA}

\author{Shun-Qing Shen}
\affiliation{Department of Physics and Centre of Theoretical and Computational Physics,
The University of Hong Kong, Pokfulam Road, Hong Kong, China}

\date{\today}

\begin{abstract}
We study the quantum diffusive transport of multivalley massive Dirac cones, where time-reversal symmetry requires opposite spin orientations in inequivalent valleys. We show that the intervalley scattering and intravalley scattering can be distinguished from the quantum conductivity that corrects the semiclassical Drude conductivity, due to their distinct symmetries and localization trends.
In immediate practice, it allows transport measurements to estimate the intervalley scattering rate in hole-doped monolayers of group-VI transition metal dichalcogenides (e.g., molybdenum dichalcogenides and tungsten dichalcogenides), an ideal class of materials for valleytronics applications. The results can be generalized to a large class of multivalley massive Dirac systems  with spin-valley coupling and time-reversal symmetry.
\end{abstract}


\maketitle
The Bloch bands in many crystals have degenerate
but inequivalent extrema in the crystal momentum space, known as valleys. Like the spin index in spintronics, the valley index is a well-defined degree of freedom for low-energy carriers and, thus, can be used to encode information as well. This has led to the concept of valleytronics, a new type of electronics based on manipulating the valley index of carriers~\cite{Gunawan06prl,Rycerz07natphys,Xiao07prl,Yao08prb,Yao09prl}. The timescale of intervalley scattering determines how long the information represented by valley polarization can be retained. For valleytronics applications to be practical, this time scale shall be long as compared to the typical time for the control of valley dynamics. It is thus crucial to measure the intervalley scattering time for identifying potential valleytronics materials, in particular, in a transport scenario.

Several extensively studied monolayer 2D crystals are promising materials to host valley-based electronics, including graphene, and graphenelike crystals such as silicene \cite{Lalmi10apl, Padova10apl, Aufray10apl, Chen12prl,Liu11prl} and monolayer group-VI transition metal dichalcogenides \cite{Novoselov05pnas,Mak10prl,WangFeng10nl,Lebegue09prb}. In these hexagonal 2D crystals, both the conduction and valence band edges are at the two inequivalent valleys at $K$ points (corners of the first Brillouin zone), which are related by time reversal. In monolayer dichalcogenides, groundbreaking theoretical and experimental progresses on the dynamical control of valleys were recently achieved~\cite{Xiao12prl,Zeng12natnano,Heinz12natnano,PKU12natcomm}.
Unlike graphene, monolayer dichalcogenides are described by massive Dirac fermions, and intrinsic spin-orbit coupling (SOC) gives rise to splitting of valence bands with opposite spins; the splitting must be opposite at the two valleys as required by time-reversal symmetry (see e.g., Fig. 1). This effective coupling between the spin and valley indices can have two significant consequences for valleytronics: (i) the interplay between spin and valley degrees of freedom and (ii) a unique form of intervalley scattering that must be accompanied by a simultaneous spin flip.

\begin{figure}[t]
\centering
\includegraphics[width=0.45\textwidth]{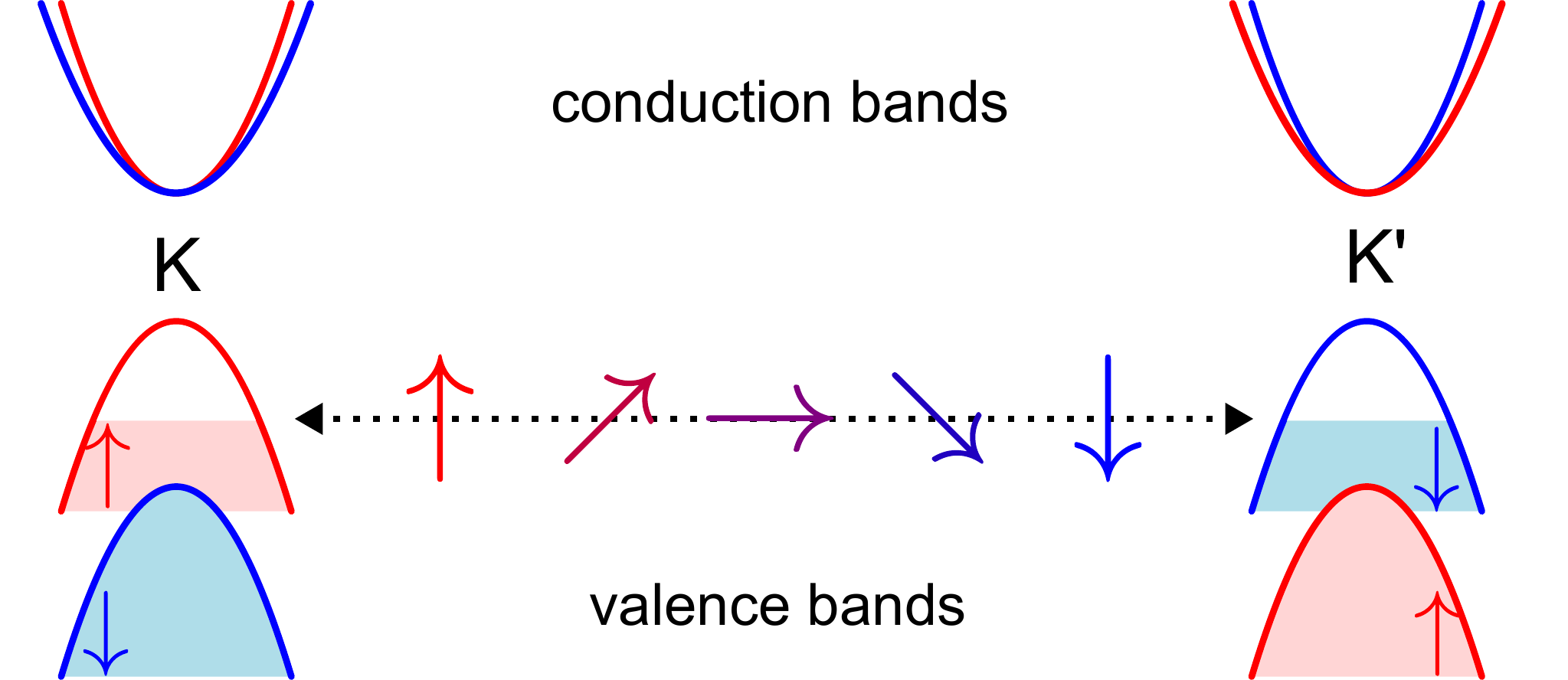}
\caption{Low-energy effective band structures of the $MX_2$ monolayer at $K$ and $K'$ valleys.
The Fermi surface (at the top edges of the shadowed areas) intersects the highest two valence bands of opposite spin orientations (marked by $\uparrow$ and $\downarrow$).
Horizontal dotted arrows indicate the intervalley spin-flip scattering. }
\label{fig:model}
\end{figure}

In this Letter, we study the quantum diffusive transport of multivalley massive Dirac fermions with spin-valley coupling. Without loss of generality, we choose the model of monolayer dichalcogenides $MX_2$ ($M$ = Mo, W; $X$
= S, Se) for a concrete discussion. This problem is theoretically unique from the conventional 2D electron gas systems or the massless graphene.
We show that the spin-valley coupled band edges result in distinct symmetries and localization behaviors in intra- and intervalley scattering dominant regimes. Namely, the spin-conserved intravalley scattering leads to negative quantum conductivity and positive magnetoconductivity from the weak localization (WL) effect, whereas the intervalley spin-flip scattering with time-reversal symmetry gives rise to positive quantum conductivity and negative magnetoconductivity from the weak antilocalization (WAL). The quantum conductivity and magnetoconductivity have logarithmic dependence on the ratio between intra- and intervalley scattering rates, making possible the measurement of the intervalley scattering rate in a vastly broad range, from 3 orders of magnitude below to 3 orders of magnitude above the intravalley scattering rate. Recently, dichalcogenide field effect transistors became experimentally accessible~\cite{Radisavljevic11natnano,Fang12nl}.
Experimental verifications of these phenomena will reveal the nature of impurities and provide general guidance on the suppression of intervalley scattering for a longer valley lifetime necessary for valleytronic applications. Our approach and results are readily to be extended to spin-valley coupled Dirac fermions with much smaller masses in monolayers of silicon, germanium, and tin \cite{Liu11prl}, as they share a similar low-energy band structure.

The band edge electrons and holes in monolayer $MX_2$ are well described by the two-valley massive Dirac model in two dimensions \cite{Xiao12prl}:
\begin{eqnarray}\label{hamiltonian}
H = \hbar v(\pm k_x\hat{\sigma}_x+k_y\hat{\sigma_y})+ \frac{\Delta}{2} \hat{\sigma}_z  \pm \lambda \hat{s}_z \otimes \frac{  \hat{1}-\hat{\sigma}_z}{2} \;,
\end{eqnarray}
where $\pm $ stands for $K$ and $K'$ valleys, respectively. The Pauli matrices $ \hat{\sigma}_{x,y,z}$ act on the pseudospin indexing the $A\equiv d_{z^2}$ and $B\equiv (d_{x^2-y^2}\pm i d_{xy})/\sqrt{2}$ orbitals, while $s_{z}=\uparrow, \downarrow$ is the $z$ component of real spin. $(k_x,k_y)$ is the wave vector measured from $K$ ($K'$) points.
The Hamiltonian is described by three parameters: the SOC strength $\lambda$, the band gap $\Delta$, and the effective velocity $v$.
The resulting band structure consists of four sets of massive Dirac cones, two in $K$ and two in $K'$ valleys.  An important feature is the large SOC splitting \cite{Xiao12prl,Zhu11prb}($2\lambda \sim$ 0.15 eV in molybdenum dichalcogenides and $2\lambda \sim$ 0.4 eV in tungsten dichalcogenides) between the spin-up and spin-down states at the valence band top.  The conduction band bottoms remain degenerate.  The band dispersion is schematically shown in Fig. \ref{fig:model}.
The model, based on a tight-binding analysis, is further confirmed by first-principles calculations \cite{Feng12prb} and well describes the optical properties of this family of materials (cf., the recent experiments on valley-dependent optical selection rule \cite{Zeng12natnano,Heinz12natnano,PKU12natcomm}). Here, this model is used in a different context to address the quantum transport phenomena, which are of relevance for valleytronics applications based on transistor or other transport devices.

Because of the large spin splitting in the valence bands, in the following we focus on the localization effect in hole-doped samples, when the Fermi surface intersects the two highest valence bands as shown in Fig. \ref{fig:model}. It is instructive to first consider two limiting regimes when either intra- or intervalley scattering dominates.

(1) Intravalley scattering dominant regime. As shown in Fig. \ref{fig:model}, the intravalley scattering within each valence band can only happen between the same spin species. Meanwhile, because of the large band gap ($\Delta \sim 1.6$ eV), the orbital pseudospin is almost fully polarized and does not play a role. The frozen pseudospin can be quantitatively seen from the Berry phase~\cite{Xiao10rmp}, which for the valence bands is given by
\begin{equation}
\phi_{v}=2\pi \frac{V+\Delta-\lambda}{2V+\Delta-\lambda}\sim 2\pi
\end{equation}
where $V$ measures the Fermi energy from the valence band top. $\Delta$ is much larger than $\lambda$ and allowed $V$ ($\in[0,2 \lambda]$ eV) in $MX_2$ \cite{Xiao12prl}. The conserved spin and peuedospin rotational symmetries, along with time-reversal symmetry in this system, belong to the orthogonal symmetry class \cite{Dyson1962}. The orthogonal symmetry always predicts WL \cite{HLN80}, which is a negative quantum interference correction to electronic conductivity and usually shows a positive logarithmic magnetoconductivity.

(2) Intervalley scattering dominant regime. The intervalley scattering in $MX_2$ must break spin-rotational symmetry because of the opposite spin splitting at the two valleys (Fig.~\ref{fig:model}). The broken spin-rotational symmetry together with time-reversal symmetry leads to the symplectic symmetry \cite{Dyson1962}. The symplectic symmetry always promises WAL \cite{HLN80}, which is a positive quantum interference correction to electronic conductivity and usually shows a negative logarithmic magnetoconductivity.

\begin{figure}[t]
\centering
\includegraphics[width=0.5\textwidth]{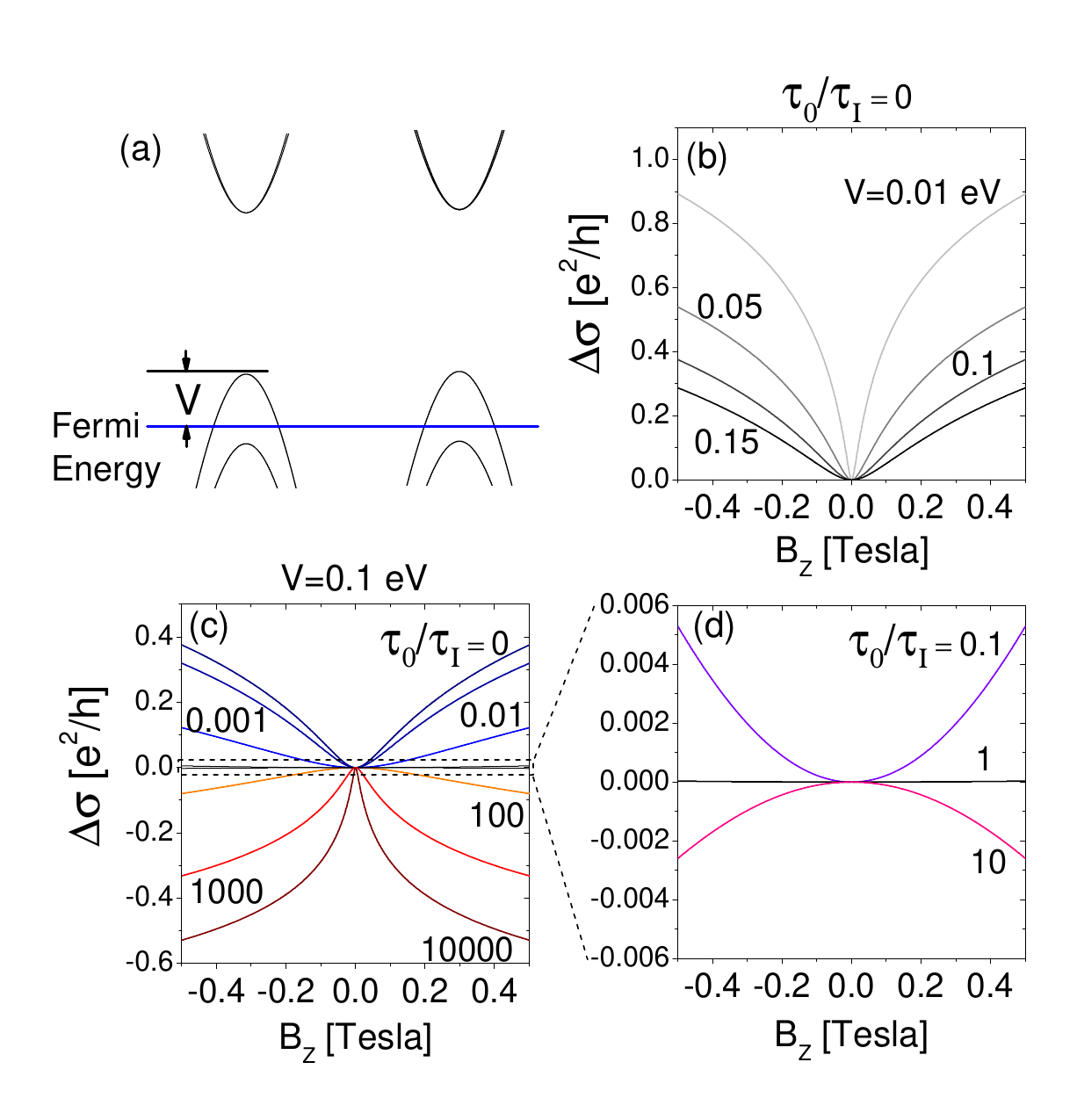}
\caption{(a) Sketch of band structure. $V$ measures the Fermi energy from the valence band top ($V\in[0,2\lambda]$eV). [(b)-(d)] Magnetoconductivity $\Delta\sigma$ when the Fermi energy intersects the highest valence bands. The positive (negative) logarithmic magnetoconductivity is the signature for the WL (WAL), which indicates that intravalley (intervalley) scattering is stronger than intervalley (intravalley) scattering. (b) Only WL in the absence of intervalley scattering ($\tau_0/\tau_{\mathrm{I}}=0$) for different $V$. (c) The crossover between WL and WAL at $V=0.1$ eV and for different $\tau_0/\tau_{\mathrm{I}}$. A larger $\tau_0/\tau_{\mathrm{I}}$ means stronger intervalley scattering. (d) Zoom-in of (c) for $\tau_0/\tau_{\mathrm{I}}=0.1,1,10$. Parameters: $\Delta=1.66$ eV and $\lambda=0.075$ eV \cite{Xiao12prl}, mean free path $\ell=10$ nm, and phase coherence length $\ell_{\phi}=300$ nm.  }
\label{fig:mc-inter}
\end{figure}

To test the above picture, we generalize the diagrammatic techniques \cite{Bergmann84PhysRep, Suzuura02prl, McCann06prl,Imura09prb,Lu11prl,Shan12prb} to these spin-valley coupled multivalley massive Dirac cones.
The method is based on expanding the Kubo formula of conductivity, with Eq.~(\ref{hamiltonian})
as the unperturbed part and scattering potentials as perturbations.
The spin-conserved intravalley scattering is modeled by
\begin{equation}
U^{0}_{\mathbf{k},\mathbf{k}'}  =
 \sum_{\mathbf{R} }u^{\mathbf{R} } e^{i(\mathbf{k}'-\mathbf{k})\cdot \mathbf{R} },
\end{equation}
where $u^{\mathbf{R}}$ is the potential of an impurity at position $\mathbf{R}$. $\mathbf{k},\mathbf{k}'$ are electron wave vectors. Although the intravalley scattering should be related to long-range potentials, the short-range potential and the delta correlation used for both $U^0$ have been justified numerically \cite{Yan08prl}. In the basis of $|+$$\uparrow$$ A\rangle $, $|+$$\uparrow$$ B\rangle$, $|-$$ \downarrow$$ A\rangle $, $|-$$ \downarrow$$ B\rangle $, the spin-flip intervalley scattering that preserves time-reversal symmetry can be in general modeled by
 \begin{eqnarray}
 U^{\mathrm{I }}_{\mathbf{k},\mathbf{k}'}  &=&
\left[
  \begin{array}{cccc}
   U^{A}_z  & 0 & U^{A}_- & 0 \\
    0 & U^{B}_z & 0 & U^{B}_- \\
    U^{A}_+ & 0 & -U^{A}_z & 0 \\
    0 & U^{B}_+ & 0 & -U^{B}_z \\
  \end{array}
\right], \nonumber\\
U^{A/B}_{z}& =& i\sum_{\mathbf{R}\in A/B }u_z^{\mathbf{R}} e^{i(\mathbf{k}'-\mathbf{k})\cdot \mathbf{R}} , \nonumber\\
U^{A/B}_{\pm} &=&  i\sum_{\mathbf{R}\in A/B}e^{i(\mathbf{k}'-\mathbf{k})\cdot \mathbf{R}} (u_x^{\mathbf{R}}+iu_y^{\mathbf{R}}),
\end{eqnarray}
where $\hat{x}, \hat{y}, \hat{z}$ are the unit vectors. The $i$ in front of $U^{\mathrm{I }}_{\mathbf{k},\mathbf{k}'}$ protects time-reversal symmetry as $i$ and spins formed by \{$|+$$\uparrow$$ A\rangle$, $|-$$ \downarrow$$ A\rangle $\} and \{$|+$$\uparrow$$ B\rangle$, $|-$$ \downarrow$$ B\rangle $\} change sign under time reversal.
For two possible spin-flip mechanisms, the magnetic scattering and spin-orbit scattering, the latter preserves time-reversal symmetry and is likely to exist in the materials. For the spin-orbit scattering $(u^{\mathbf{R}}_x,u^{\mathbf{R}}_y,u^{\mathbf{R}}_z)=u^{\mathbf{R}}\mathbf{k}\times \mathbf{k}'\cdot( \hat{x},\hat{y}, \hat{z})$, where $\mathbf{k}$ and $\mathbf{k}'$ also change sign under time reversal.

\begin{figure}[t]
\centering
\includegraphics[width=0.45\textwidth]{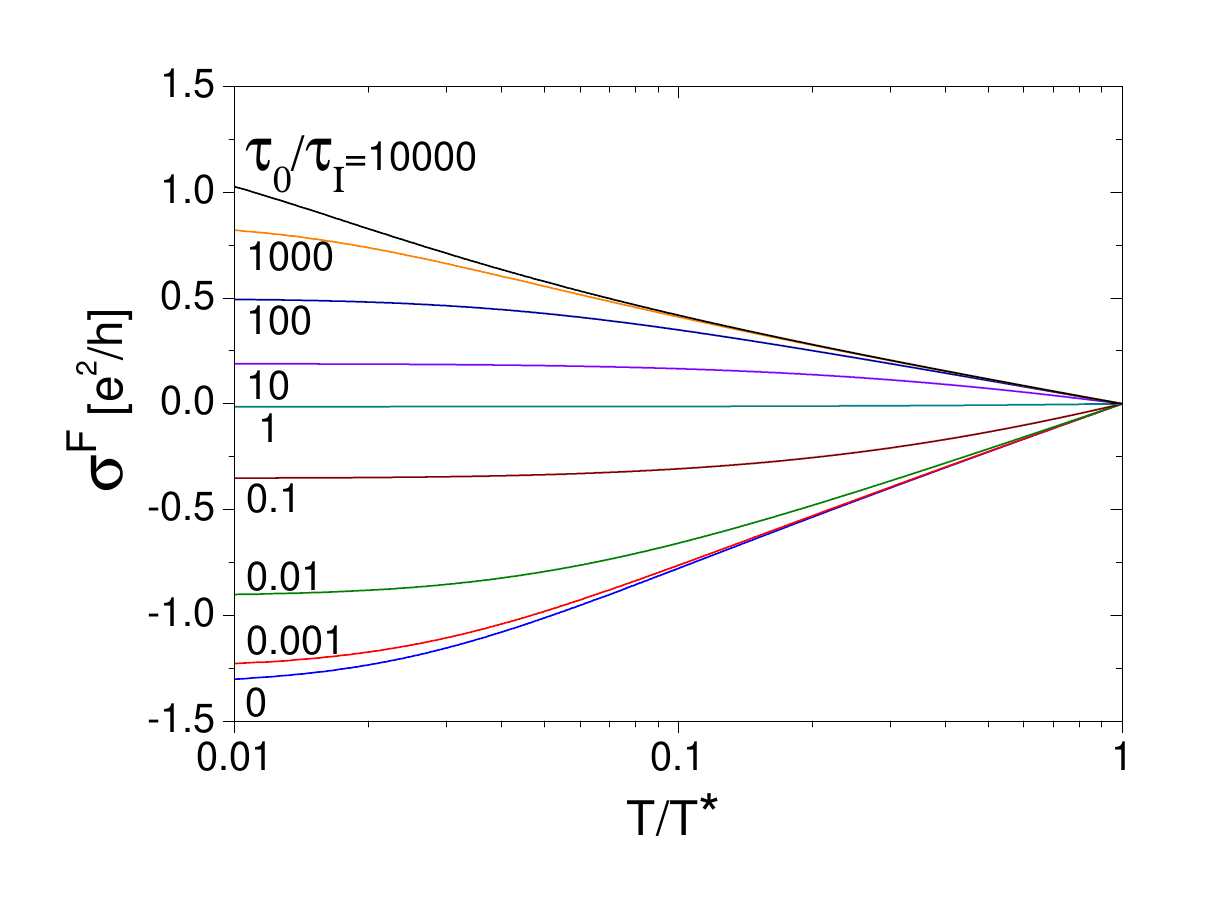}
\caption{The quantum conductivity $\sigma^F$ as a function of temperature $T$ for different $\tau_0/\tau_{\mathrm{I}}$.
Positive (negative) $\sigma^F$ corresponds to WAL (WL).
Other parameters are the same as those in Figs. \ref{fig:mc-inter} (c) and (d). The signature that intervalley (intravalley) scattering is stronger than intravalley (intervalley) scattering is given by a positive (negative) $\sigma^F$, as well as by the fact that $\sigma^F$ increases (decreases) with decreasing $T$.}
\label{fig:temperature}
\end{figure}

The total conductivity is given by
\begin{eqnarray}
\sigma=\sigma^D+\sigma^F,
\end{eqnarray}
where $\sigma^D$ is the semiclassical (Drude) conductivity following the Einstein relation
\begin{eqnarray}
\sigma^D =  e^2   N_F D
\end{eqnarray}
where $N_F=E_F/(\pi \hbar^2 v^2)$ is the two-valley density of states of at the Fermi energy $E_F$ measured from the Dirac point, $D= \eta_v v_F^2 \tau /2 $ is the diffusion constant, $\eta_v$ is the ladder diagram correction to the velocity of Dirac fermions \cite{Shon98jpsj}, $v_F$ is the Fermi velocity, and the total scattering time is given by $\tau=(1/\tau_0+1/\tau_{\mathrm{I}})^{-1}$.
$\tau_{0}$ and $\tau_{\mathrm{I}}$ are intravalley and intervalley scattering times, respectively. Since $\sigma^D$ is a function of the total scattering time, it cannot distinguish the contributions from intervalley and intravalley scattering. Besides, $\sigma^D$ is insensitive to magnetic field.

The quantum interference correction to the Drude conductivity (quantum conductivity, for short) is found in forms of logarithmic functions
\begin{eqnarray}\label{sigma_F}
\sigma^{F}= \frac{e^2 }{\pi h} \left( C_{0} \ln\frac{X_{0}^{-2}+\ell_{\phi}^{-2} }{X_{0}^{-2}+\ell^{-2} }+ C_{\mathrm{I}} \ln\frac{X_{\mathrm{I}}^{-2}+\ell_{\phi}^{-2} }{X_{\mathrm{I}}^{-2}+\ell^{-2} }\right),
\end{eqnarray}
where $C_\mathrm{0}$ and $C_{\mathrm{I}}$ are weight factors for the intravalley and intervalley contributions.
$X_0$ and $X_{\mathrm{I}}$ are corresponding characteristic lengths that effectively reduce the mean free path $\ell$ and phase coherence length $\ell_{\phi}$.
The intervalley/intravalley scattering ratio is incorporated in the expressions for $C_{0,\mathrm{I}}$ and $X_{0,\mathrm{I}}$ \cite{sup}.
Due to its interference origin, $\sigma^F$ can be suppressed by a perpendicular magnetic field $B_z$, giving rise to the magnetoconductivity
\begin{equation}\label{MC}
\Delta \sigma (B_z) \equiv \sigma^F(B_z)-\sigma^F(0)  =\frac{e^{2}}{\pi h}(C_{\mathrm{0}} F_{\mathrm{0}} + C_{\mathrm{I}}F_{\mathrm{I}}),
\end{equation}
where $F_i= \Psi (\ell_{B}^{2}/\ell _{i}^{2}+\frac{1}{2})-\ln ( \ell _{B}^{2}/\ell_{i }^{2})$, $1/\ell_{i}^2$ $\equiv$ $1/\ell_{\phi}^2+1/X^2_i$, $\Psi $ is the digamma function, and $\ell _{B}\equiv \sqrt{\hbar/(4e|B_z|)} $ is the magnetic length \cite{sup}.
Considering the low mobility in $MX_2$ \cite{Radisavljevic11natnano}, we have assumed short $\tau$ and that $\ell\equiv \sqrt{D\tau}$; then, the results are not sensitive to $\ell$ and $\tau$.

Figure~\ref{fig:mc-inter}(b) shows the magnetoconductivity without intervalley scattering when the Fermi energy intersects with the two highest valence bands. The WL is a quantum interference induced suppression of conductivity; it can be lifted by magnetic field and gives positive magnetoconductivity. Figure~\ref{fig:mc-inter}(b) shows that the magnetoconductivity is always positive, corresponding to the WL of the single valence band in the allowed range of the Fermi energy. Figures~\ref{fig:mc-inter}(c) and (d) show the magnetoconductivity as the intervalley scattering increases, where the ratio between scattering times
\begin{eqnarray}
 \tau_0/\tau_{\mathrm{I}}
\end{eqnarray}
increases with increasing intervalley scattering.
As $\tau_0/\tau_{\mathrm{I}}$ increases, the magnetoconductivity in Figs.~\ref{fig:mc-inter}(c) and (d) changes from positive to negative, corresponding to a crossover from WL to its opposite, the WAL.
The magnetoconductivity covers a wide range of $\tau_0/\tau_{\mathrm{I}}$, from $<0.001$ to $>1000$. When $\tau_0/\tau_{\mathrm{I}} > 10 $ or $<0.1$, a small change in the magnetoconductivity corresponds to a large change of $\tau_0/\tau_{\mathrm{I}}$.
This sensitivity allows us to estimate even a very small $\tau_{\mathrm{I}}$ when $\tau_0$ dominates, or vice versa.
Note that a completely opposite crossover happens in graphene, where intervalley scattering leads to WL while intravalley scattering gives WAL \cite{Suzuura02prl,Tikhonenko09prl}, because graphene is gapless and has ignorable spin-orbit interaction \cite{Min06prb,Yao07prb}.

The signature of the crossover is also provided by the sign of the quantum conductivity $\sigma^F$ and how $\sigma^F$ changes with decreasing temperature. $\sigma^F$ becomes nonzero at low temperatures when the phase coherence length $\ell_{\phi}$ is longer than the mean free path $\ell$. $\ell_{\phi}$ increases as the decoherence mechanisms from electron-phonon and electron-electron interactions are suppressed at low temperatures.
Empirically, $\ell_{\phi}=C_{ph}  (T/T^*)^{-3/2}+C_{ee}(T/T^*)^{-1/2}$, where the coefficients $C_{ph}$ and $C_{ee}$ are for electron-phonon and electron-electron interactions, respectively. $T^*$ is a characteristic temperature at which $\ell_{\phi}=\ell$ and $\sigma^F$ vanishes, so it defines the boundary between the quantum and classical diffusion regimes. Usually, $T^*$ ranges between 10 and 100 K \cite{Ando82rmp,Lee85rmp,Sarma11rmp}. In the calculation, we assume that $\ell_{\phi}=300$ nm at $T/T^*=0.01$ and $\ell_{\phi}= \ell$ at $T/ T^*=1$.
Figure~\ref{fig:temperature} shows $\sigma^F(T)$ in Eq.~(\ref{sigma_F}) for different $\tau_0/\tau_{\mathrm{I}}$.
The WL (WAL) can be read from a negative (positive) $\sigma^F$, as well as the fact that $\sigma^F$ decreases (increases) with decreasing temperature.
Note that the total conductivity is $\sigma(T)=\sigma^D+\sigma^F(T,B)$, where the Drude conductivity $\sigma^D$ may give an extra temperature power law that depends on scattering mechanisms \cite{Ando82rmp,Sarma11rmp}. But $\sigma^F $ from the weak (anti)localization is the contribution most sensitive to magnetic field (because of the logarithmic dependence). $\sigma^F$ thus can be extracted from the total conductivity by $\sigma^F(T)=\sigma(T,0)-\sigma(T,B_c)$, with $B_c$ a finite magnetic field strong enough to quench the weak (anti)localization of, say, about 1 Tesla.

\begin{figure}[t]
\centering
\includegraphics[width=0.4\textwidth]{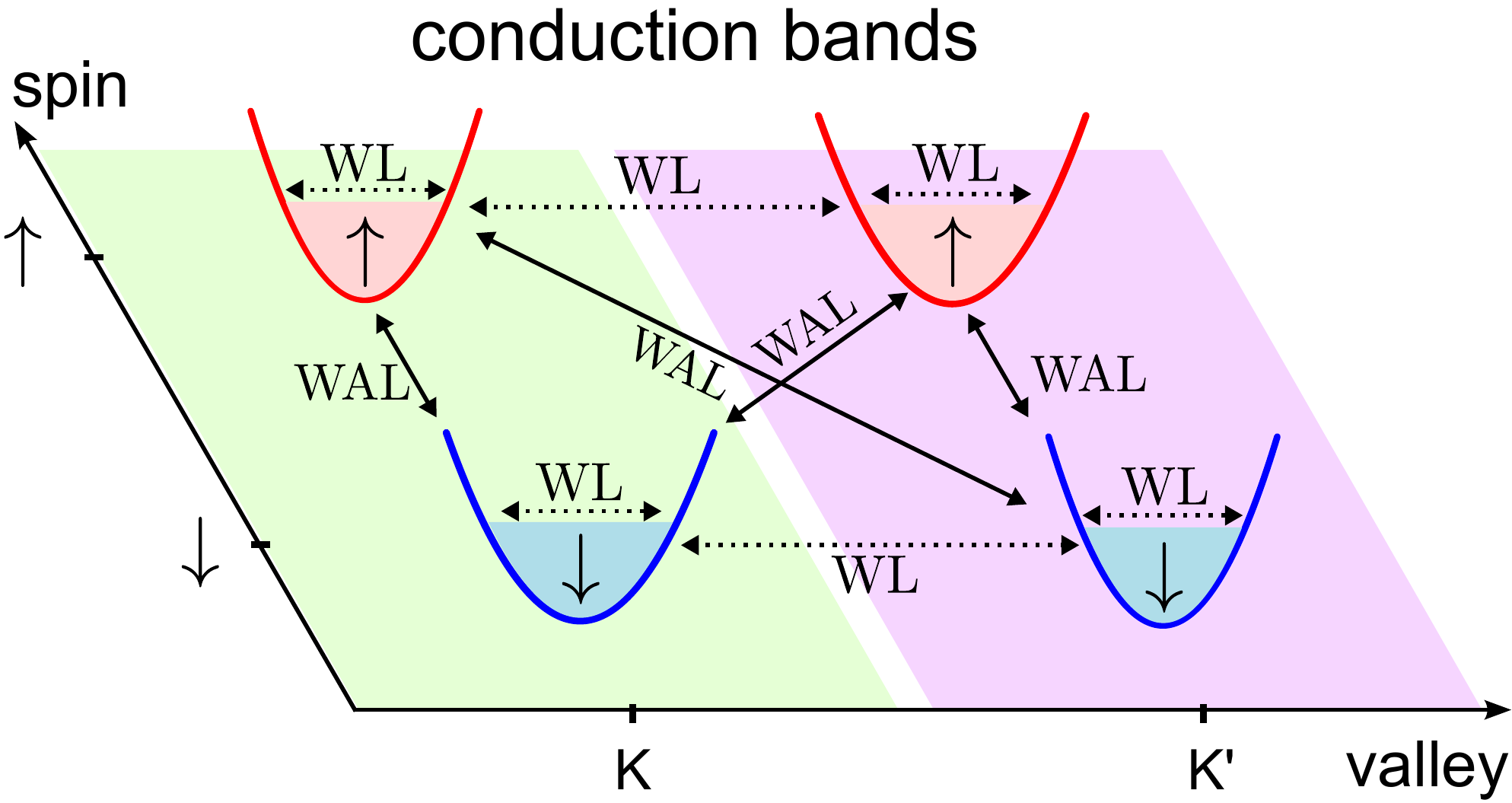}
\caption{WL and WAL in the conduction bands. Spin-up ($\uparrow$) and spin-down ($\downarrow$) are offset for clarity. The solid and dotted arrows represent spin-flip and spin-conserved scattering, respectively. }
\label{fig:conduction}
\end{figure}

We can also have some qualitative arguments for the conduction bands.
All four conduction bands can take part in transport as they always intersect the Fermi surface together. Therefore, different from the transport in hole-doped samples, both intravalley spin-flip and intervalley spin-conserved scattering are possible in the conduction bands. With time-reversal symmetry, the spin-flip scattering leads to WAL, while spin-conserved scattering to WL, as the pseudo-spin degree of freedom is always frozen.
All types of scattering and the resulting WL and WAL are shown in Fig.~\ref{fig:conduction}.

The existing experiments on the $MX_2$ monolayers \cite{Radisavljevic11natnano,Fang12nl} are mainly focused on the on-off characteristics of field-effect transistors, implying that the resistance in the samples still comes mainly from the Schottky barriers between mismatched metal electrodes and monolayers, instead of from the monolayers themselves. We expect our theory to inspire more experimental efforts for realizing good Ohm contacts (e.g., by liquid gating) and to explore the spin-valley coupled physics in transport.

This work was supported by the Research Grant Council
of Hong Kong under Grant No. HKU 705110P (H.Z.L. and S.Q.S.) and HKU 706412P (W.Y.). D.X. was supported by the U.S. Department of Energy, Office of Basic Energy Sciences, Materials Sciences and Engineering Division.

\appendix

\section{Supplemental material}

\subsection{Model}

The effective Hamiltonian of$MX_2$ reads \cite{Xiao12prl}
\begin{eqnarray}
H = \hbar v(\pm k_x\hat{\sigma}_x+k_y\hat{\sigma_y})+\frac{\Delta}{2}\hat{\sigma}_z \pm \lambda \hat{s}_z \otimes \frac{\hat{1}-\hat{\sigma}_z}{2}
\end{eqnarray}
where $\pm $ for $K$ and $K'$ valleys, respectively. The Pauli matrices $ \sigma_{x,y,z}$ act on the psudospin from its $A\equiv d_{z^2}$ and $B\equiv (d_{x^2-y^2}\pm i d_{xy})/\sqrt{2}$ orbitals while $s_{z}=\uparrow, \downarrow$ is the $z$-component of real spin. $(k_x,k_y)$ is the wave vector measured from $K$ ($K'$) points.
$\lambda$ depicts the spin-orbit splitting of valence bands. $\Delta$ is the gap without spin-orbit splitting. $\hbar$ is Planck's constant over $2\pi$. $v$ is the effective Fermi velocity.
The Hamiltonian describes four massive Dirac cones, two in $K$ and two in $K'$ valleys. Each cone contains a pair of conduction and valence bands, with their dispersions given by
\begin{eqnarray}
E^{+,\uparrow}_{c/v} =E^{-,\downarrow}_{c/v}&=&  \frac{\lambda}{2}\pm\sqrt{\frac{(\Delta-\lambda)^2}{4}+(\hbar v k)^2} \\
E^{+,\downarrow}_{c/v}=E^{-,\uparrow}_{c/v} &=&  -\frac{\lambda}{2}\pm\sqrt{\frac{(\Delta+\lambda)^2}{4}+(\hbar v k)^2}
\end{eqnarray}
with $c/v$ for conduction and valence bands, respectively. $k^2=k_x^2+k_y^2$. An important feature is the splitting ($2\lambda\sim$ 0.15 eV) between the valence bands with spin $\uparrow$ and $\downarrow$ orientations. The dispersions are schematically shown in Fig. 1 in the manuscript for $K$ and $K'$ valleys.
We are particularly interested in the situation when the Fermi surface intersects two highest valence bands.
Their eigen wavefunctions are written as
\begin{eqnarray}
|+ \uparrow  v\rangle  =
 \left[
                   \begin{array}{c}
                     1 \\
                     0 \\
                   \end{array}
                 \right] \otimes \left[
                   \begin{array}{c}
                     b \\
                     -a e^{i\varphi} \\
                   \end{array}
                 \right] ,\
|-  \downarrow  v\rangle  =
\left[
                   \begin{array}{c}
                   0\\
                   1\\
                     \end{array}
                 \right]\otimes\left[
                   \begin{array}{c}
                     b e^{i\varphi}\\
                     a  \\
                   \end{array}
                 \right]\nonumber
\end{eqnarray}
where $a=\cos\frac{\theta}{2}$, $b=\sin\frac{\theta}{2}$, $
\cos\theta  = (\Delta-\lambda)/ (2V+\Delta-\lambda)$, and $V$ measures the Fermi energy from the top of the valence bands.
Here and after, we work in the subspace spanned by the basis functions in the order of $|+$$\uparrow$$ A\rangle $, $|+$$\uparrow$$ B\rangle$, $|-$$ \downarrow$$ A\rangle $, $|-$$ \downarrow$$ B\rangle $.

\subsection{Impurity potentials}

Because spin is a good quantum number in each band,
we consider ordinary elastic intravalley scattering in each band, with the potential modeled by
\begin{equation}
U^{0}(\mathbf{r})=\sum_{\mathbf{R}} u(\mathbf{r}-\mathbf{R}),
\end{equation}
where $u(\mathbf{r}-\mathbf{R})$ is the potential of an impurity at position $\mathbf{R}$.
For spin-flip scattering that also preserves time-reversal symmetry, it is natural to consider spin-orbit scattering
\begin{eqnarray}
U^{\mathrm{I}}(\mathbf{r})= \sum_{\mathbf{R}} \nabla u(\mathbf{r}-\mathbf{R})\times \mathbf{k} \cdot  \mathbf{s},
\end{eqnarray}
where $\nabla$ gives the gradient of $u$. The Pauli matrices $\mathbf{s}$ operate on the spin basis functions from
$\{ |+$$\uparrow$$A\rangle,  |-$$\downarrow$$A\rangle \}$ or from $\{|+$$\uparrow$$B\rangle, |-$$\downarrow$$B\rangle\}$.
We assume that $u(\mathbf{r}-\mathbf{R})=u^{\mathbf{R}}\delta(\mathbf{r}-\mathbf{R})$, $\langle U(\mathbf{r})\rangle_{\mathrm{imp}
}=0$, and $\langle U(\mathbf{r})U(\mathbf{r^{\prime }})\rangle _{\mathrm{imp}}\sim \delta (\mathbf{r}-\mathbf{r}^{\prime })$, where $U=U^0+U^{\mathrm{I}}$ and $\langle ...\rangle _{\mathrm{imp}}$ means average over impurity configurations.
Although intravalley scattering should be related to long-range potential, the practice by the delta potential is justified by the numerical calculations \cite{Yan08prl}.

Between two plane waves
\begin{eqnarray}
|\mathbf{k}\rangle \equiv   e^{i\mathbf{k}\cdot \mathbf{r}}/\sqrt{S}, \ \
|\mathbf{k}'\rangle \equiv e^{i\mathbf{k}'\cdot \mathbf{r}}/\sqrt{S},
\end{eqnarray}
where $S$ is the area, the scattering matrix elements of $U^{0}(\mathbf{r})$ and $U^{\mathrm{I},\mathrm{so}}(\mathbf{r})$ are found as
\begin{eqnarray}
U^{0}_{\mathbf{k},\mathbf{k}'}  =
 \sum_{\mathbf{R} }u^{\mathbf{R} } e^{i(\mathbf{k}'-\mathbf{k})\cdot \mathbf{R} }
\end{eqnarray}
and
\begin{eqnarray}
 U^{\mathrm{I}}_{\mathbf{k},\mathbf{k}'}  =
\left[
  \begin{array}{cccc}
   U^{A}_z  & 0 & U^{A}_- & 0 \\
    0 & U^{B}_z & 0 & U^{B}_- \\
    U^{A}_+ & 0 & -U^{A}_z & 0 \\
    0 & U^{B}_+ & 0 & -U^{B}_z \\
  \end{array}
\right],
\end{eqnarray}
where
\begin{eqnarray}
U^A_{z} &=& i\sum_{\mathbf{R} }u^{\mathbf{R} } e^{i(\mathbf{k}'-\mathbf{k})\cdot \mathbf{R} } \mathbf{k}\times \mathbf{k}'\cdot \hat{z}, \nonumber\\
U^A_{\pm} &=& i\sum_{\mathbf{R} }u^{\mathbf{R} }e^{i(\mathbf{k}'-\mathbf{k})\cdot \mathbf{R} } \mathbf{k}\times \mathbf{k}' \cdot (\hat{x} \pm i \hat{y} ),
\end{eqnarray}
$\hat{x}, \hat{y}, \hat{z}$ stand for the unit vectors,
and $A\rightarrow B$ for $U^B_{z,\pm}$. The extra $i$ in $U^{\mathrm{I}}_{\mathbf{k},\mathbf{k}'}$ comes from the integration by parts
\begin{eqnarray}
\int d\mathbf{r} e^{i\mathbf{k}\cdot \mathbf{r}}\nabla u^{\mathbf{R}}\delta(\mathbf{r}-\mathbf{R})  = - \int d\mathbf{r}u^{\mathbf{R}}\delta(\mathbf{r}-\mathbf{R})\nabla e^{i\mathbf{k}\cdot\mathbf{r}}
\end{eqnarray}
and protects time-reversal symmetry in spin-orbit scattering as $i$, spin, momenta change sign under time reversal.

The total scattering rate is given by
\begin{eqnarray}
\tau^{-1} &=& \tau_{0}^{-1}+ \tau_{\mathrm{I}}^{-1},
\end{eqnarray}
where the intravalley ordinary scattering rate
\begin{eqnarray}
\tau_{0}^{-1}
&\equiv &\frac{2\pi}{\hbar} \sum_{\mathbf{k}'} \langle | \langle + \uparrow v |U^{0}_{\mathbf{k} ,  \mathbf{k}'   } | + \uparrow v \rangle  |^2 \rangle_{\mathrm{imp}}
\delta(E_F - E^{+\uparrow}_{v\mathbf{k}'}) \nonumber\\
&=&  \frac{2\pi}{\hbar } N V_0 (a^4 + b^4)
\end{eqnarray}
and the intervalley scattering rate
\begin{eqnarray}
\tau_{\mathrm{I}}^{-1} &\equiv &\frac{2\pi}{\hbar} \sum_{\mathbf{k}'}  \langle | \langle + \uparrow v |U^{\mathrm{I}}_{ \mathbf{k} \mathbf{k}'} | - \downarrow v \rangle |^2 \rangle_{\mathrm{imp}}
\delta(E_F - E^{-\downarrow }_{v \mathbf{k}'})\nonumber\\
&=&
\frac{2\pi}{\hbar } N (b^4 V^A_{\mathrm{I}} + a^4 V^B_{\mathrm{I}}) \overline{(\mathbf{k}\times \mathbf{k}')^2},
\end{eqnarray}
$N$ is the density of states at the Fermi level $E_F$ per spin per valley, $(\mathbf{k}\times \mathbf{k}')^2$ is replaced by its average $\overline{(\mathbf{k}\times \mathbf{k}')^2}$ \cite{HLN80}, and
\begin{eqnarray}
V_0 \equiv \langle \sum_{\mathbf{R}} \sum_{\mathbf{R}'} u^{\mathbf{R}} u^{\mathbf{R}'}
e^{i(\mathbf{k}-\mathbf{k}')\cdot (\mathbf{R}-\mathbf{R}')} \rangle_{\mathrm{imp}},
\end{eqnarray}
\begin{eqnarray}
V^A_{\mathrm{I}} \equiv 2\langle \sum_{\mathbf{R} } \sum_{\mathbf{R}'} u^{\mathbf{R} } u^{\mathbf{R}'}
e^{i(\mathbf{k}-\mathbf{k}')\cdot (\mathbf{R}-\mathbf{R}')} \rangle_{\mathrm{imp}}
\end{eqnarray}
represent correlations between scattering, $A\rightarrow B$ for $V^B_{\mathrm{I}} $.
We follow the practical assumption that scattering are correlated only when they are of the same type and from the same orbital \cite{McCann06prl,Suzuura02prl}.
To measure the ratio of intervalley scattering rate to the total scattering rate, we define that
\begin{eqnarray}
\eta_{\mathrm{I}}\equiv  \frac{\tau^{-1}_{\mathrm{I}}}{\tau_0^{-1}+\tau_{\mathrm{I}}^{-1}}= \frac{1}{1+(\tau_0/\tau_{\mathrm{I}})^{-1}}.
\end{eqnarray}

\subsection{Spin-orbit scattering}

The spin-orbit interaction near an impurity at $\mathbf{R}$ is given by
\begin{eqnarray}
U_{\mathbf{R}}(\mathbf{r})=\frac{1}{4m^2c^2}   \nabla \tilde{u}(\mathbf{r}-\mathbf{R})\times \mathbf{p} \cdot  \mathbf{s}
\end{eqnarray}
The total potential of spin-orbit scattering then can be written as a summation over all the impurities
\begin{eqnarray}
U(\mathbf{r})= \sum_{\mathbf{R}} \nabla u(\mathbf{r}-\mathbf{R})\times \mathbf{k} \cdot  \mathbf{s}
\end{eqnarray}
where $u$ absorbs $\tilde{u}$ and other parameters. The value of $U(\mathbf{r})$ between two plane waves of momenta $\mathbf{k}$ and $\mathbf{k}'$, known as the Born amplitude, is given by
\begin{eqnarray}
U_{\mathbf{k},\mathbf{k}' } &\equiv &\frac{1}{S}\int d\mathbf{r} e^{-i\mathbf{k}\cdot \mathbf{r}} U(\mathbf{r}) e^{i\mathbf{k}'\cdot \mathbf{r}} \nonumber\\
& = & \frac{1}{S} \sum_{\mathbf{R}} \int d\mathbf{r} e^{i(\mathbf{k}'-\mathbf{k})\cdot \mathbf{r}} \nabla u(\mathbf{r}-\mathbf{R})   \times \mathbf{k} \cdot  \mathbf{s}.
\end{eqnarray}
Integration by parts gives an extra $i$ when acting $\nabla$ on the plane-wave part
\begin{eqnarray}
U_{\mathbf{k},\mathbf{k}' }
 =  i\frac{1}{S} \sum_{\mathbf{R}} \int d\mathbf{r}  \ u(\mathbf{r}-\mathbf{R})  e^{i(\mathbf{k}'-\mathbf{k})\cdot \mathbf{r}}
\mathbf{k}\times \mathbf{k}' \cdot  \mathbf{s}.
\end{eqnarray}
Assuming $u(\mathbf{r}-\mathbf{R}) =  u^{\mathbf{R}}\delta(\mathbf{r}-\mathbf{R}) $
\begin{eqnarray}
U_{\mathbf{k},\mathbf{k}' }
& = & i \sum_{\mathbf{R}}u^{\mathbf{R}} e^{i(\mathbf{k}'-\mathbf{k})\cdot \mathbf{R}}
\mathbf{k}\times \mathbf{k}' \cdot  \mathbf{s}
\end{eqnarray}
where $u^{\mathbf{R}}$ is random in both amplitude and sign at different $\mathbf{R}$, so $\langle u^{\mathbf{R}}\rangle_{\mathrm{imp}}=0$. Besides, we assume that $\langle u^{\mathbf{R}}u^{\mathbf{R}'}\rangle_{\mathrm{imp}}\propto \delta(\mathbf{R}-\mathbf{R}')$.

\subsection{Quantum conductivity from intervalley scattering }

The weak localization and antilocalization come from the quantum interference correction to conductivity, and can be calculated from a summation of the diagrams known as Hikami boxes \cite{Suzuura02prl,McCann06prl}, denoted as
\begin{eqnarray}
\sigma ^{F}=\sigma^{++}_{++}+\sigma^{--}_{--}+\sigma^{+-}_{-+}+\sigma^{-+}_{+-},
\end{eqnarray}
where $+$ is short for $|+\uparrow v\rangle \otimes|\mathbf{k}\rangle $ and $-$ for $|- \downarrow v \rangle \otimes|\mathbf{k}'=\mathbf{q}-\mathbf{k}\rangle$.
$\mathbf{q}$ is the summation
of momenta before and after scattering.
Scattering from $\mathbf{k}$ to $\mathbf{q}-\mathbf{k}$ describes backscattering as $\mathbf{q}\rightarrow 0$.
$\sigma^{++}_{++}$ and $\sigma^{--}_{--}$ from one valley of massive Dirac cone have been found \cite{Lu11prl}. $\sigma^{-+}_{+-}$ and $\sigma^{+-}_{-+}$ from intervalley scattering will be calculated here. We have
\begin{eqnarray}\label{Deltasigmaxx1_rest}
\sigma^{+-}_{-+}&=&\frac{e^2  \hbar}{2\pi}\sum_{\mathbf{k}}\widetilde{v}^x_{+}\widetilde{v}^x_{-}
G^R_{+}G^A_{+} G^R_{-}G^A_{-} \sum_{\mathbf{q}} \Gamma^{+-}_{-+} ,
\end{eqnarray}
where
\begin{eqnarray}
\widetilde{v}^x_{\pm}=\frac{\eta_v}{\hbar}\frac{\partial E_{\pm}}{\partial k_x}
\end{eqnarray}
is the dressed velocity, $\eta_v$ can be calculated from the Ladder diagrams \cite{Shon98jpsj}. The Green functions are given by
\begin{eqnarray}
G^{R}_{\pm } =(G^{A}_{\pm})^*=  \frac{1}{E_F-E_{\pm}+ i \hbar/2\tau}.
\end{eqnarray}
$E_F$ is the Fermi energy.
The vertex function $\Gamma^{+-}_{-+}$ satisfy a set of Bethe-Salpeter equations
\begin{eqnarray}
\Gamma^{+-}_{-+} &=& \gamma^{+-}_{-+} + \sum_{\mathbf{k}_{\mu}}\sum_{\nu=\pm} \gamma^{+\nu}_{-\overline{\nu}} G^{R}_{\mathbf{k}_{\mu},\nu} G^A_{\mathbf{q}-\mathbf{k}_{\mu},\overline{\nu}} \Gamma^{\nu-}_{\overline{\nu}+}\nonumber\\
\Gamma^{\nu-}_{\overline{\nu}+} &=& \gamma^{\nu-}_{\overline{\nu}+}
+ \sum_{\mathbf{k}_{\mu}}\sum_{\nu'=\pm}
\gamma^{\nu\nu'}_{\overline{\nu}\overline{\nu}'}
G^{R}_{\mathbf{k}_{\mu},\nu'} G^A_{\mathbf{q}-\mathbf{k}_{\mu},\overline{\nu}'} \Gamma^{\nu'-}_{\overline{\nu}'+},\nonumber
\end{eqnarray}
where $\overline{\nu}$=$-\nu$.
The bare vertex is found as
\begin{eqnarray}
\gamma^{+ - }_{- + }   &\equiv &
\langle \langle - \downarrow  v | U^{\mathrm{I},\mathrm{so}}_{\mathbf{k} ,\mathbf{k}'} | + \uparrow  v\rangle
\langle + \uparrow  v|U^{\mathrm{I},\mathrm{so}}_{-\mathbf{k},-\mathbf{k}'}  | -  \downarrow  v\rangle  \rangle_{\mathrm{imp}}\nonumber\\
&=& \frac{\hbar}{2\pi N \tau } \mathbf{a}_{11}  e^{i(\varphi'-\varphi)} ,\ \ \ \mathbf{a}_{11}=\eta_I,
\end{eqnarray}
where $\varphi$ and $\varphi'$ are momentum angles of $\mathbf{k}$ and $\mathbf{k}'=\mathbf{q}-\mathbf{k}$. It is fair to ignore $\gamma^{++}_{--}$ because
scattering in one valley should be hardly correlated to those in the other.
By assuming the ansatz for the full vertex functions
\begin{eqnarray}
\Gamma^{+ -}_{- +}\equiv  \frac{\hbar}{2\pi N \tau}\sum_{n,m\in\{0,1,2\}}\mathbf{A}_{nm}e^{i(n\varphi'-m\varphi )},
\end{eqnarray}
the coupled Bethe-Salpeter equations reduce to one for the coefficient matrices
\begin{eqnarray}
\mathbf{A}
&=&\mathbf{a}+\mathbf{a}\mathbf{\Phi} \mathbf{a}\mathbf{\Phi} \mathbf{A},
\end{eqnarray}
where the Kernel from $\sum_{\mathbf{k}_{\mu}}G^{R}_{\mathbf{k}_{\mu},\nu} G^A_{\mathbf{q}-\mathbf{k}_{\mu},\overline{\nu}}$ is found as
\begin{eqnarray}
\mathbf{\Phi}
=\left[
  \begin{array}{ccc}
    1 -\frac{1}{2}Q^2 &  \frac{i}{2}  Q_+ & -\frac{1}{4} Q_+^2 \\
      \frac{i}{2}  Q_- & 1-\frac{1}{2}Q^2 &  \frac{i}{2} Q_+ \\
     -\frac{1}{4} Q_-^2 &   \frac{i}{2} Q_- & 1-\frac{1}{2}Q^2 \\
  \end{array}
\right]\nonumber
\end{eqnarray}
with $Q_{\pm}=(q_x\pm iq_y)v\tau\sin\theta $ and $Q^2=Q_+Q_-$. The matrix equation gives
\begin{eqnarray}
\Gamma^{+- }_{- +} \approx   -\frac{\hbar}{2\pi N \tau }\frac{\eta_{\mathrm{I}}}{1-\eta_{\mathrm{I}}^2+Q^2 \eta_{\mathrm{I}}^2},
\end{eqnarray}
where the minus sign comes from $e^{i(\varphi-\varphi')}\approx e^{i\pi}$ in the small $\mathbf{q}$ limit.
With $\Gamma^{+- }_{- +}$ and
\begin{eqnarray}
\sum_{\mathbf{k}} \widetilde{v}^x_{+}\widetilde{v}^x_{-}
 G^R_{+}G^A_{+} G^R_{-}G^A_{-} = -2\pi N(\tau/\hbar)^3\eta_v^2 v^2\sin^2\theta,
\end{eqnarray}
Eq. (\ref{Deltasigmaxx1_rest}) becomes
\begin{eqnarray}\label{13}
\sigma^{+-}_{-+}
=  \frac{e^2 }{h} \frac{\eta_v^2 }{\eta_{\mathrm{I}} } \sum_{\mathbf{q}} \frac{1}{X_{\mathrm{I}}^{-2} +q^2 }, \ \
X_{\mathrm{I}}^{-2} = \frac{1/\eta_{\mathrm{I}}^2-1}{ 2 \ell^2 \sin^2\theta}
\end{eqnarray}
where
\begin{eqnarray}
\eta_{v} = \frac{1}{1-(1-\eta_{\mathrm{I}}) \frac{a^2b^2}{a^4+b^4}},
\end{eqnarray}
and the mean free path is defined as $\ell\equiv v\tau/\sqrt{ 2} $.
If there is no intervalley scattering, i.e., $\eta_{\mathrm{I}}\rightarrow 0$, $\sigma^{+-}_{-+}$ vanishes.
If the intervalley scattering overwhelms the intravalley scattering, i.e., $\eta_{\mathrm{I}} \rightarrow 1$, the intervalley Cooperon gap $X_{\mathrm{I}}^{-2}$ vanishes, then $\sigma^{+-}_{-+}$ becomes divergent as $q\rightarrow 0$ and dominates the total $\sigma^F$. $\sigma^{-+}_{+-}=\sigma^{+-}_{-+}$ can be found similarly. Performing the integrals over $q$ in Eq. (\ref{13}) between the inverses of the phase coherence length $\ell_{\phi}^{-1}$ and mean free path $\ell^{-1}$ gives $\sigma^{+-}_{-+}$, the total zero-field quantum conductivity from both intravalley (`` $\mathrm{0}$ ") and intervalley (`` $\mathrm{I}$ ") scattering is found
\begin{eqnarray}\label{sigma_F-app}
\sigma^{F}= \frac{e^2 }{\pi h} \left( C_{0} \ln\frac{X_{0}^{-2}+\ell_{\phi}^{-2} }{X_{0}^{-2}+\ell^{-2} }+ C_{\mathrm{I}} \ln\frac{X_{\mathrm{I}}^{-2}+\ell_{\phi}^{-2} }{X_{\mathrm{I}}^{-2}+\ell^{-2} }\right),
\end{eqnarray}
where
\begin{eqnarray}
C_\mathrm{I}&=&-\frac{\eta_v^2}{2\eta_I} \nonumber\\
C_{\mathrm{0}}&=&\frac{\eta _{v}^{2}(\frac{1}{2}+\eta _{H})(g_1-g_0)^2}{(g_1+g_0+1)(g_1-g_0)-g_0^2/g_2},
\end{eqnarray}
\begin{eqnarray}
X_{\mathrm{0}}^{-2}= \frac{g_0g_1(g_1-g_0)}{(2 \ell^2\sin^2\theta)
[(g_1+g_0+1)(g_1-g_0)-g_0^2/g_2]},
\end{eqnarray}
\begin{eqnarray}
g_{0}= 2(\frac{a^4+b^4}{a^4}\frac{1}{1-\eta_{\mathrm{I}}}-1),\nonumber\\
g_{1}=2(\frac{a^4+b^4}{2a^2b^2}\frac{1}{1-\eta_{\mathrm{I}}}-1), \nonumber\\
g_{2}=2(\frac{a^4+b^4}{b^4}\frac{1}{1-\eta_{\mathrm{I}}}-1).
\end{eqnarray}
$\eta_{H}=-(1-\eta_{\mathrm{I}})a^2b^2/(a^4+b^4)/2$ comes from the dressed Hikami boxes \cite{McCann06prl}.
Above only the bare Hikami box of $\sigma^{+-}_{-+}$ is calculated,
the dressed Hikami boxes can be approximated by multiplying $1+2\eta_H$ but this correction can be ignored in the large-gap limit where $\eta_H\rightarrow 0$ \cite{Lu11prl}.

Under a perpendicular magnetic field $B_z$, $q^{2}$ in Eq. (\ref{13}) will be quantized into
\begin{eqnarray}
q_{n}^{2}=(n+1/2)\frac{4eB_z}{\hbar } \equiv (n+1/2)/\ell _{B}^{2},
\end{eqnarray}
where $n$ labels the Landau levels, $\ell _{B}\equiv \sqrt{\hbar/(4e|B_z|)} $ is the magnetic length. Summation over $n$ gives the quantum conductivity $\sigma
^{F}(B_z)$ at finite field \cite{Bergmann84PhysRep}. The total magnetoconductivity (when $\ell\ll \ell_B$) is found as
\begin{equation}\label{MC}
\Delta \sigma (B_z) \equiv
\sigma ^{F}(B_z)-\sigma ^{F} =\frac{e^{2}}{\pi h}(C_{\mathrm{0}} F_{\mathrm{0}} + C_{\mathrm{I}}F_{\mathrm{I}}),
\end{equation}
where $F_i= \Psi ( \ell_{B}^{2}/\ell _{i}^{2}+\frac{1}{2})-\ln ( \ell _{B}^{2}/\ell_{i }^{2})$, $1/\ell_{i}^2$ $\equiv$ $1/\ell_{\phi}^2+1/X^2_i$, and $\Psi $ is the digamma function.

\end{document}